\documentstyle[11pt,amssymb,epsfig,psfig]{article}

\begin{document}

\title{ Casimir energy-momentum tensor for a brane in de Sitter spacetime }

\author{
A. A. Saharian  $^{1}$\footnote{E-mail: saharyan@server.physdep.r.am
}
and  M. R. Setare $^2$ \footnote{E-mail: rezakord@ipm.ir} \\
 {$^1$ Department of Physics, Yerevan
State University, Yerevan, Armenia } \\
and
\\ {$^2$ Physics Department, Institute for Studies in Theoretical Physics and }
\\ {Mathematics, Tehran, Iran} \\ {Department of Science, Physics Group, Kurdistan University, Sanandeg, Iran}}

\maketitle

\begin{abstract}
Vacuum expectation values of the
energy-momentum tensor for a conformally coupled scalar field is investigated in de Sitter (dS) spacetime in presence of a curved brane on which the field obeys the Robin boundary condition with coordinate dependent coefficients. To generate the corresponding vacuum densities we use the conformal relation between dS and Rindler spacetimes and the results previously obtained by one of the authors for the Rindler counterpart. The resulting energy-momentum tensor is non-diagonal and induces anisotropic vacuum stresses. The asymptotic behaviour of this tensor is investigated near the dS horizon and the boundary.
\end{abstract}

\bigskip

{PACS number(s): 03.70.+k, 11.10.Kk}

\newpage

\section{Introduction} \label{sec:int}

De Sitter (dS) spacetime is the maximally symmetric solution of Einsten's equation with a positive cosmological constant. Recent astronomical observations of supernovae and cosmic microwave background \cite{Ries98} indicate that the universe is accelerating and can be well approximated by a world with a positive cosmological constant. If the universe would accelerate indefinitely, the standard cosmology leads to an asymptotic dS universe. De Sitter spacetime plays an important role in the inflationary scenario, where an exponentially expanding approximately dS spacetime is employed to solve a number of problems in standard cosmology. The quantum field theory on dS spacetime is also of considerable interest. In particular, the inhomogeneities generated by fluctuations of a quantum field during inflation provide an attractive mechanism for the structure formation in the universe. Another motivation for investigations of dS based quantum theories is related to the recently proposed holographic duality between quantum gravity on dS spacetime and a quantum field theory living on boundary identified with the timelike infinity of dS spacetime \cite{Stro01}.

The one of most striking macroscopic manifestations of quantum properties is the Casimir effect. The presence of reflecting boundaries alters the zero-point modes of a quantized field, and results in the shifts in the vacuum
expectation values of quantities quadratic in the field, such as
the energy density and stresses. In particular, vacuum forces
arise acting on constraining boundaries. The particular features
of these forces depend on the nature of the quantum field, the
type of spacetime manifold, the boundary
geometries and the specific boundary conditions imposed on the
field. Since the original work by Casimir in 1948 \cite{Casi48}
many theoretical and experimental works have been done on this
problem (see, e.g., \cite{Most97,Plun86,Lamo99,Bord99,Bord01,Kirs01,Bord02,Milt02,Romeo,Elizalde} and
references therein). Many different approaches have been used: mode summation method, Green function formalism, multiple scattering expansions, heat-kernel series, zeta function regularization technique, etc. Recently new methods are developed for the Casimir energy calculations in given background fields \cite{Bord96,Grah02,Gies03}.

The Casimir effect can be viewed as a polarization of vacuum by boundary conditions. The interaction of fluctuating quantum fields with background gravitational fields give rise to another type of vacuum polarization (see, for instance, \cite{Birrell,Grib94}). Here we will study an example where both types of polarizations are present. Namely, we evaluate the vacuum expectation values for the enrgy-momentum tensor of a conformally coupled scalar field on background of $D+1$-dimensional dS spacetime when a curved brane is present (for investigations of the Casimir energy in braneworld models with dS branes see Refs. \cite{Noji00,Nayl02,Eliz03}). As a brane we take $D$-dimensional hypersurface which is the conformal image of a plate moving with constant proper acceleration in the Rindler stacetime. We will assume that the field is prepared in the state conformally related to the Fulling--Rindler vacuum in the Rindler spacetime. To generate the vacuum expectation values in dS bulk, we use the conformal relation between dS and Rindler spacetimes and the results from \cite{Saha02} for the corresponding Rindler problem with mixed boundary conditions. Previously this method has been used in \cite{set5} to derive the vacuum stress on parallel plates for a scalar field with Dirichlet boundary conditions in de Sitter spactime and in Ref. \cite{set6}
to investigate the vacuum characteristics of the Casimir configuration
on background of conformally flat brane-world geometries for
massless scalar field with Robin boundary conditions on plates.

The present paper is organized as follows. In the next section the geometry of our problem and the conformal relation between dS and Rindler spacetimes are discussed. The results are presented for the vacuum expectation values of the energy-momentum tensor for a scalar field induced by a plate uniformly accelerated through the Fulling--Rindler vacuum. In Section \ref{sec2}, by using the formula relating the renormalized energy-momentum tensors for conformally related problems in combination with the appropriate coordinate transformation, we derive expressions for the vacuum energy-momentum tensor in dS space. The main results are rementioned and summarized in Section \ref{secconc}.

\section{Conformal relation between dS and Rindler problems} \label{sec1}

Consider a conformally coupled massless scalar field $%
\varphi (x)$ satisfying the equation
\begin{equation}
\left( \nabla _{l}\nabla ^{l}+\zeta R\right) \varphi
(x)=0,\quad \zeta =\frac{D-1}{4D} , \label{fieldeq}
\end{equation}
on background of a $D+1$--dimensional dS spacetime. In Eq.
(\ref{fieldeq}), $\nabla _{l}$ is the operator of the covariant
derivative, and $R$ is the Ricci scalar for the corresponding
metric $g_{ik}$. In static coordinates $x^i=(t,r,\theta ,\theta
_2,\ldots ,\theta _{D-2},\phi )$ dS metric has the form
\begin{equation}
ds_{{\rm dS}}^{2}=g_{ik}dx^idx^k=\left( 1-\frac{r^{2}}{\alpha
^{2}}\right) dt^{2}-\frac{dr^{2}}{1-\frac{r^{2}}{\alpha
^{2}}}-r^{2}d\Omega ^{2}_{D-1} , \label{ds2dS}
\end{equation}
where $d\Omega ^{2}_{D-1}$ is the line element on the
$D-1$--dimensional unit sphere in Euclidean space, and the
parameter $\alpha $ defines the dS curvature radius. Note that
$R=D(D-1)/\alpha ^2$. We will assume that the field satisfies the
mixed boundary condition
\begin{equation}
\left( A+Bn^{l}\nabla _{l}\right) \varphi (x)=0,\quad x\in S ,
\label{boundcond}
\end{equation}
on the hypersurface $S$, $n^{l}$ is the normal to this surface,
$n_{l}n^{l}=-1$ (the form of the hypersurface will be specified below, see Eq. (\ref{hypersurf})). The results in the following will depend on the
ratio of Robin coefficients $A$ and $B$. However, to keep the transition
to the Dirichlet and Neumann cases transparent we will use the
form (\ref{boundcond}). Our main interest in the present paper is
to investigate the vacuum
expectation value (VEV) of the energy--momentum tensor for the field $%
\varphi (x)$ induced by the hypersurface $S$. The presence of
boundaries modifies the spectrum of the zero--point fluctuations
compared to the case without boundaries and results in the shift
in the VEV's of physical quantities, such as vacuum energy
density and stresses. This is the well known Casimir effect.

\begin{figure}[tbph]
\begin{center}
\epsfig{figure=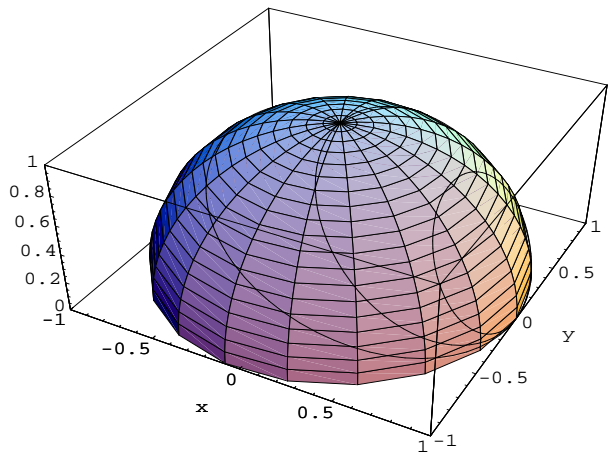,width=8cm,height=6cm}
\end{center}
\caption{ The section of dS spacetime for fixed $(t,\theta
_2,\ldots ,\theta _{D-2},\phi )$ embedded into $3D$ Euclidean space
with coordinates $(x,y,z)=(\frac{r}{\alpha}\cos \theta
,\frac{r}{\alpha}\sin \theta ,z)$ and described by equation
(\ref{surf2}). The curves on this surface correspond to the
boundary (\ref{hypersurf}) with  $a_{\alpha }=0.5,1,2$. }
\label{fig1surf}
\end{figure}

To make maximum use of the flat spacetime calculations, first of all let us present the dS line element in the form
conformally related to the Rindler metric. With this aim we
make the coordinate transformation $x^i\to x'^{i}=(\tau ,\xi
,{\mathbf{x}}' )$, ${\mathbf{x}}'=(x'^{2},\ldots ,x'^{D})$ (see
Ref. \cite{Birrell} for the case $D=3$)
\begin{eqnarray}
&& \tau =\frac{t}{\alpha },\quad \xi =\frac{\sqrt{\alpha ^2-r^2}
}{\Omega },\quad x'^{2}=\frac{r}{\Omega }\sin \theta \cos \theta
_2, \ldots ,\nonumber \\
&&  x'^{D-2}=\frac{r}{\Omega }\sin \theta \sin \theta _2\ldots
\sin \theta _{D-3}\cos \theta _{D-2}, \label{coord} \\
&& x'^{D-1}=\frac{r}{\Omega }\sin \theta \sin \theta _2\ldots \sin
\theta _{D-2}\cos \phi, \quad x'^{D}=\frac{r}{\Omega }\sin \theta
\sin \theta _2\ldots \sin \theta _{D-2}\sin \phi, \nonumber
\end{eqnarray}
with the notation
\begin{equation}\label{Omega}
  \Omega =1-\frac{r}{\alpha }\cos \theta .
\end{equation}
Under this coordinate transformation the dS line element takes the
form
\begin{equation}
ds_{{\rm dS}}^{2}=g'_{ik}dx'^idx'^k=\Omega ^2\left( \xi ^{2}d\tau
^{2}-d\xi ^{2}-d{\mathbf{x}}'^{2}\right) .  \label{ds2dS1}
\end{equation}
In this form the dS metric is manifestly conformally related to
the Rindler spacetime with the line element $ds_{{\rm R}}^{2}$:
\begin{equation}
ds_{{\rm dS}}^{2}=\Omega ^{2}ds_{{\rm R}}^{2},\quad ds_{{\rm
R}}^{2}=g^{{\mathrm{R}}}_{ik}dx'^idx'^k=\xi ^{2}d\tau ^{2}-d\xi
^{2}-d{\mathbf{x}}'^{2},\quad g'_{ik}=\Omega ^2
g^{{\mathrm{R}}}_{ik}. \label{confrel}
\end{equation}
By using the standard transformation formula for the vacuum
expectation values of the energy-momentum tensor in conformally
related problems (see, for instance, \cite{Birrell}), we can
generate the results for dS spacetime from the corresponding
results for the Rindler spacetime. In this paper as a Rindler
counterpart we will take the the vacuum energy-momentum tensor
induced by an infinite plate moving by uniform proper acceleration
through the Fulling--Rindler vacuum. We will assume that the plate
is located in the right Rindler wedge and has the coordinate $\xi
=a$. Observe that in coordinates $x^i$ the boundary $\xi =a$ is presented by
the hypersurface
\begin{equation}\label{hypersurf}
  \sqrt{\alpha ^2-r^2}=a\left( 1-\frac{r}{\alpha }\cos \theta \right)
\end{equation}
in dS spacetime. As a boundary $S$ in Eq. (\ref{boundcond}) we will take this hypersurface. In Fig. \ref{fig1surf} we have plotted the
section of dS spacetime for fixed $(t,\theta _2,\ldots ,\theta
_{D-2},\phi )$. The corresponding surface is embedded into $3D$
Euclidean space with coordinates $(x,y,z)=(\frac{r}{\alpha}\cos
\theta ,\frac{r}{\alpha}\sin \theta ,z)$ and is defined by the
equation
\begin{equation}\label{surf2}
x^2+y^2+z^2=1,\quad z\geq 0.
\end{equation}
In coordinates $(x,y,z)$ the boundary (\ref{hypersurf}) is defined
by the intersection of the surface (\ref{surf2}) with the cylinder
\begin{equation}\label{surf3}
  \left( x-\frac{a_{\alpha }^2}{1+a_{\alpha }^2}\right) ^2+
  \frac{y^2}{1+a_{\alpha }^2}=\frac{1}{(1+a_{\alpha }^2)^2},
  \quad a_{\alpha }=\frac{a}{\alpha }.
\end{equation}
The corresponding curves on the surface (\ref{surf2}) are plotted
in Fig. \ref{fig1surf} for values $a_{\alpha }=0.5,1,2$. In the
limit $a_{\alpha }\to 0$ the curves tend to the dS horizon
presented by the circle $(r/\alpha =1,z=0)$. For all avlues of $a_{\alpha }$ the hypersurfaces (\ref{surf3}) touch the dS horizon at $(r,\theta )=(\alpha ,0)$.

The expectation values of the energy-momentum tensor induced by
the presence of an infinite plane boundary moving with uniform
acceleration through the Fulling-Rindler vacuum were investigated
by Candelas and Deutsch \cite{Cand77} for a conformally coupled
$4D$ Dirichlet and Neumann massless scalar and electromagnetic
fields. In this paper the region of the right Rindler wedge to the
right of the barrier is considered. In Ref. \cite{Saha02} we have
investigated the Wightman function and the VEV of the energy-momentum tensor for a massive scalar field with general curvature coupling parameter, satisfying the Robin
boundary conditions on an infinite plane in an arbitrary number
of spacetime dimensions. Both regions, including the one between the barrier and Rindler horizon, are considered. Recently, the total Casimir energy in this problem is investigated \cite{Saha03R} by using the zeta function regularization technique. The expectation values of the energy-momentum
tensor for a scalar field $\varphi _{{\mathrm{R}}}(x')$ in the
Fulling--Rindler vacuum can be presented in the form of the sum
\begin{equation}
\left\langle 0_{R}|T_{i}^{k}[g^{{\mathrm{R}}}_{lm},\varphi
_{{\mathrm{R}}}]|0_{R}\right\rangle =\left\langle
\tilde{0}_{R}|T_{i}^{k}[g^{{\mathrm{R}}}_{lm},\varphi
_{{\mathrm{R}}}]|\tilde{0}_{R}\right\rangle +\left\langle
T_{i}^{k}[g^{{\mathrm{R}}}_{lm},\varphi _{{\mathrm{R}}}]\right\rangle
^{(b)}, \label{TikR}
\end{equation}
where $|0_{R}\rangle $ and $|\tilde 0_{R}\rangle $ are the
amplitudes for the vacuum states in the Rindler space in presence and
absence of the plate respectively, and $\left\langle
T_{i}^{k}[g^{{\mathrm{R}}}_{lm},\varphi _{{\mathrm{R}}}]\right\rangle
^{(b)}$ is the part of the vacuum energy-momentum tensor induced
by the plate. Note that the state $|\tilde 0_{R}\rangle $ corresponds to the standard Fulling--Rindler vacuum. In the case of a conformally coupled massless scalar
field, for the part without boundaries one has (see Ref. \cite{Cand77} for the case $D=3$ and Ref. \cite{Saha02} for an arbitrary $D$)
\begin{equation}
\left\langle \tilde{0}_{R}|T_{i}^{k}[g^{{\mathrm{R}}}_{lm},\varphi
_{{\mathrm{R}}}]|\tilde{0}_{R}\right\rangle = \frac{a_D\xi
^{-D-1}}{2^{D-1}\pi ^{D/2}\Gamma (D/2)}{\rm diag}\left(
-1,\frac{1}{D},\ldots ,\frac{1}{D}\right) , \label{TikR0}
\end{equation}
with the notation
\begin{equation}\label{aD}
a_D=\int_{0}^{\infty }\frac{\omega ^{D}d\omega }{e^{2\pi \omega
}+(-1)^{D}}\prod_{l=1}^{l_{m}}\left[ \left( \frac{D-1-2l}{2\omega
}\right) ^{2}+1\right],
\end{equation}
where $l_{m}=D/2-1$ for even $D>2$ and $l_{m}=(D-1)/2$ for odd
$D>1$, and the value for the product over $l$ is equal to 1 for
$D=1,2,3$. For a scalar field $\varphi _{{\mathrm{R}}}(x')$,
satisfying the mixed boundary condition
\begin{equation}\label{boundRind}
\left( A_{{\mathrm{R}}}+B_{{\mathrm{R}}}n'^l_{{\mathrm{R}}}\nabla
'_{l} \right) \varphi _{{\mathrm{R}}}(x')=0,\quad \xi =a,\quad
n'^l_{{\mathrm{R}}}=\delta ^l_1,
\end{equation}
with constants $A_{{\mathrm{R}}}$, $B_{{\mathrm{R}}}$, the
boundary induced part in the region $\xi >a$ is defined by the
formula \cite{Saha02}
\begin{equation}
\left\langle T_{i}^{k}[g^{{\mathrm{R}}}_{lm},\varphi
_{{\mathrm{R}}}]\right\rangle ^{(b)}=\frac{-\delta
_{i}^{k}}{2^{D-2}\pi ^{(D+1)/2}\Gamma \left( \frac{D-1}{2}\right)
}\int_{0}^{\infty
}dkk^{D-1}\int_{0}^{\infty }d\omega \frac{\bar{I}_{\omega }(ka)}{%
\bar{K}_{\omega }(ka)}F^{(i)}[K_{\omega }(k\xi )],  \label{TikRb}
\end{equation}
where the functions $F^{(i)}[g(z)]$ for $i=0,1$ have the form
\begin{eqnarray}
F^{(0)}[g(z)] &=&\left( \frac{1}{2}-2\zeta \right) \left( \frac{dg(z)}{dz%
}\right) ^{2}+\frac{\zeta }{z}\frac{d}{dz} g^2(z)+\left[ \frac{1}{2}-2\zeta +\left( \frac{1}{2}+2\zeta \right)
\frac{\omega
^{2}}{z^{2}}\right] g^2(z),  \label{Figz} \\
F^{(1)}[g(z)] &=&-\frac{1}{2}\left( \frac{dg(z)}{dz}\right)
^{2}-\frac{\zeta }{z}\frac{d}{dz}g^2(z)+\frac{1}{2}\left( 1-\frac{\omega ^{2}}{z^{2}}\right) g^2(z),  \nonumber
\end{eqnarray}
and the functions $F^{(i)}[g(z)]$ for $i=2,\ldots ,D$ are determined by the
zero trace condition for the energy-momentum tensor,
\begin{equation}
F^{(i)}[g(z)]=-\frac{1}{D-1}\left\{ F^{(0)}[g(z)]+F^{(1)}[g(z)]\right\}
,\quad i=2,\ldots ,D.  \label{Figz1}
\end{equation}
In Eq. (\ref{TikRb}), $I_{\omega }(z)$ and $K_{\omega} (z)$ are
the Bessel modified functions and for a given function $f(z)$ we
use the notation
\begin{equation}\label{barnot}
\bar f(z)=A_{{\mathrm{R}}}f(z)+B_{{\mathrm{R}}}zf'(z).
\end{equation}
The expression for the boundary part of the vacuum energy-momentum
tensor in the region $\xi <a$ is obtained from formula
(\ref{TikRb}) by replacements $I_{\omega}\to K_{\omega}$,
$K_{\omega}\to I_{\omega}$.

\section{Vacuum energy-momentum tensor in dS bulk}
\label{sec2}

To find the VEV's induced by the surface
(\ref{hypersurf}) in dS spacetime, first we will consider the
corresponding quantities in the coordinates $(\tau ,\xi
,{\mathbf{x}}')$ with metric (\ref{ds2dS1}). These quantities can be
found from the corresponding results in the Rindler spacetime by using the standard transformation formula for the
conformally related problems \cite{Birrell}:
\begin{equation}
\left\langle 0_{{\mathrm{dS}}}|T_{i}^{k}\left[ g'_{lm},\varphi
\right] |0_{{\mathrm{dS}}}\right\rangle =\Omega
^{-D-1}\left\langle 0_{{\mathrm{R}}}|T_{i}^{k}\left[
g^{{\mathrm{R}}}_{lm},\varphi _{{\mathrm{R}}}\right]
|0_{{\mathrm{R}}}\right\rangle +\left\langle T_{i}^{k}\left[
g'_{lm},\varphi \right] \right\rangle ^{(an)},
\label{conftransemt}
\end{equation}
where the second summand on the right is determined by the trace
anomaly and is related to the divergent part of the corresponding
effective action:
\begin{equation}
\left\langle T_{i}^{k}\left[ g'_{lm},\varphi \right]
\right\rangle ^{(an)}=2g'^{kl}\frac{\delta }{\delta g'^{il}(x)}W_{{\rm %
div}}[g'_{mn},\varphi ].  \label{gravemt}
\end{equation}
Note that in odd spacetime dimensions the conformal anomaly is
absent and the corresponding anomaly part vanishes:
\begin{equation}
\left\langle T_{i}^{k}\left[ g'_{lm},\varphi \right] \right\rangle
^{(an)}=0,\quad {\rm for\;even}\;D.  \label{gravemteven}
\end{equation}
For an odd number of spatial dimensions the anomaly part in dS apacetime has the form
\begin{equation}\label{anompart1}
\left\langle T_{i}^{k}\left[ g'_{lm},\varphi \right] \right\rangle
^{(an)}=\frac{b_D}{\alpha ^{D+1}}\delta ^{i}_k,
\end{equation}
with the numerical coefficient $b_D$. In $D=3$ one has $b_3=1/960
\pi ^2$ \cite{Birrell}.

The formulae given above allow us to present the dS VEV's in the decomposed form similar to Eq. (\ref{TikR}):
\begin{equation}
\left\langle 0_{{\rm dS}}|T_{i}^{k}\left[ g'_{lm},\varphi
\right]|0_{{\rm dS}}\right\rangle =\left\langle \tilde{0}_{{\rm
dS}}|T_{i}^{k}\left[ g'_{lm},\varphi \right]|\tilde{0}_{{\rm
dS}}\right\rangle +\left\langle T_{i}^{k}\left[ g'_{lm},\varphi
\right]\right\rangle ^{(b)} , \label{TikdS}
\end{equation}
where $\left\langle \tilde{0}_{{\rm dS}}|T_{i}^{k}\left[
g'_{lm},\varphi \right]|\tilde{0}_{{\rm dS}}\right\rangle $ is
the expectation value in dS spacetime without boundaries
and the part $\left\langle T_{i}^{k}\left[ g'_{lm},\varphi
\right]\right\rangle ^{(b)}$ is induced by the hypersurface
(\ref{hypersurf}). Conformally transforming the Rindler results
one finds
\begin{eqnarray}
\left\langle \tilde{0}_{{\rm dS}}|T_{i}^{k}\left[ g'_{lm},\varphi
\right]|\tilde{0}_{{\rm dS} }\right\rangle  &=&\Omega ^{-
D-1}\left\langle \tilde{0}_{{\mathrm{R}}}|T_{i}^{k}\left[
g^{{\mathrm{R}}}_{lm},\varphi _{{\mathrm{R}}}\right]|\tilde{0}
_{{\mathrm{R}}}\right\rangle +\left\langle T_{i}^{k}\left[
g'_{lm},\varphi \right] \right\rangle ^{(an)},  \label{TikdS0} \\
\left\langle T_{i}^{k}\left[ g'_{lm},\varphi \right]\right\rangle
^{(b)} &=&\Omega ^{-D-1}\left\langle
T_{i}^{k}[g^{{\mathrm{R}}}_{lm},\varphi _{{\mathrm{R}}}]\right\rangle
^{(b)}. \label{TikdSb}
\end{eqnarray}
Under the conformal transformation $g'_{ik}=\Omega ^{2}g^{{\mathrm{R}}}
_{ik}$, the $\varphi _{{\mathrm{R}}}$ field will
change by the rule
\begin{equation}
\varphi (x')=\Omega ^{(1-D)/2}\varphi _{{\mathrm{R}}}(x'),
\label{phicontr}
\end{equation}
where the conformal factor is given by expression (\ref{Omega}).
Now by comparing boundary conditions (\ref{boundRind}) and
(\ref{boundcond}) and taking into account Eq. (\ref{phicontr}), one
obtains the relation between the coefficients in the boundary
conditions:
\begin{equation}\label{relcoef}
A=\frac{1}{\Omega }\left(
A_{{\mathrm{R}}}+\frac{D-1}{2}B_{{\mathrm{R}}} n^l\nabla _l \Omega
\right) , \quad B=B_{{\mathrm{R}}}, \quad x\in S.
\end{equation}
To evaluate the expression $n^l\nabla _l \Omega $ we need the
components of the normal to $S$ in coordinates $x^i$. They can be
found by transforming the components $n'^l=\delta ^{l}_1/\Omega $ in
coordinates $x'^{i}$:
\begin{equation}\label{normal}
n^l=\left( 0,\frac{a}{\alpha }(\cos \theta -r/\alpha ), -\frac{a }{\alpha r }\sin \theta ,0,\ldots ,0 \right) .
\end{equation}
Now it can be easily seen that $n^l\nabla _l \Omega =-\sqrt{\alpha
^2-r^2}/\alpha ^2$ and, hence, the relation between the Robin
coefficients in the Rindler and dS problems takes the form
\begin{equation}\label{relcoef1}
A=\frac{aA_{{\mathrm{R}}}}{\sqrt{\alpha
^2-r^2}}-\frac{D-1}{2}\frac{aB_{{\mathrm{R}}}}{\alpha ^2}, \quad B=B_{{\mathrm{R}}} .
\end{equation}
Note that the Robin coefficient $A$ depends on the point of the hypersurface.

The VEV's of the energy-momentum tensor in
coordinates $x^i$ with line element (\ref{ds2dS}) are obtained from expressions (\ref{TikdS0}) and
(\ref{TikdSb}) by the standard coordinate transformation formulae.
As before, we will present the corresponding components in the
form of the sum of purely dS and boundary induced parts:
\begin{equation}
\left\langle 0_{{\rm dS}}|T_{i}^{k}\left[ g_{lm},\varphi
\right]|0_{{\rm dS}}\right\rangle =\left\langle \tilde{0}_{{\rm
dS}}|T_{i}^{k}\left[ g_{lm},\varphi \right]|\tilde{0}_{{\rm
dS}}\right\rangle +\left\langle T_{i}^{k}\left[ g_{lm},\varphi
\right]\right\rangle ^{(b)} . \label{TikdS1}
\end{equation}
By using relations (\ref{coord}) between the coordinates, for
the purely dS part one finds
\begin{equation}\label{Tik0dSst}
\left\langle \tilde{0}_{{\rm dS}}|T_{i}^{k}\left[ g_{lm},\varphi
\right]|\tilde{0}_{{\rm dS}}\right\rangle = \frac{(\alpha
^2-r^2)^{-\frac{D+1}{2}}a_D}{2^{D-1}\pi ^{D/2}\Gamma
(D/2)}{\mathrm{diag}}\left( -1,\frac{1}{D},\ldots ,\frac{1}{D}
\right) +\frac{b_D}{\alpha ^{D+1}}\delta _{i}^{k}.
\end{equation}
This formula generalizes the result for $D=3$ given, for instance,
in Ref. \cite{Birrell}. As for the boundary induced
energy-momentum tensor the spatial part is anisotropic, the
corresponding part in coordinates $x^i$ is more complicated:
\begin{eqnarray}
\left\langle T_{i}^{k}\left[ g_{lm},\varphi \right]\right\rangle
^{(b)} &=& \Omega ^{-D-1}\left\langle T_{i}^{k}\left[
g^{{\mathrm{R}}}_{lm},\varphi _{{\mathrm{R}}}\right]\right \rangle
^{(b)}, \quad i,k=0,3,\ldots ,D, \label{tikv1}\\
\left\langle T_{1}^{1}\left[ g_{lm},\varphi \right]\right\rangle
^{(b)} &=&\frac{(\cos \theta -r/\alpha )^2}{\Omega
^{D+3}}\left\langle T_{1}^{1}\left[ g^{{\mathrm{R}}}_{lm},\varphi
_{{\mathrm{R}}}\right]\right\rangle ^{(b)} \nonumber \\
&& +\frac{1-r^2/\alpha ^2}{\Omega ^{D+3}} \sin ^2\theta
\left\langle T_{2}^{2}\left[ g^{\mathrm{R}}_{lm},\varphi
_{{\mathrm{R}}}\right]\right\rangle ^{(b)} , \label{tikv2}\\
\left\langle T_{1}^{2}\left[ g_{lm},\varphi \right]\right\rangle
^{(b)} &=&\frac{(r/\alpha -\cos \theta )\sin \theta }{r\Omega
^{D+3}}\left\{ \left\langle T_{1}^{1}\left[
g^{{\mathrm{R}}}_{lm},\varphi _{{\mathrm{R}}}\right]\right\rangle
^{(b)}-\left\langle T_{2}^{2}\left[ g^{{\mathrm{R}}}_{lm},\varphi
_{{\mathrm{R}}}\right]\right\rangle ^{(b)}\right\} ,\label{tikv3}\\
\left\langle T_{2}^{2}\left[ g_{lm},\varphi \right]\right\rangle
^{(b)} &=&  \frac{ 1-r^2/\alpha ^2}{\Omega ^{D+3}} \sin ^2\theta
\left\langle T_{1}^{1}\left[ g^{{\mathrm{R}}}_{lm},\varphi
_{{\mathrm{R}}}\right]\right\rangle ^{(b)}\nonumber \\
&& + \frac{(r/\alpha -\cos \theta  )^2}{\Omega ^{D+3}}
\left\langle T_{2}^{2}\left[ g^{{\mathrm{R}}}_{lm},\varphi
_{{\mathrm{R}}}\right]\right\rangle ^{(b)} ,\label{tikv4}
\end{eqnarray}
where the expressions for the components of the boundary induced energy-momentum tensor in the Rindler spacetime are given by formulae (\ref{TikRb})--(\ref{Figz1}) in the region $\xi >a$ and by similar formulae with the replacements $I_{\omega }\to K_{\omega }$ and $K_{\omega }\to I_{\omega }$ in the region $\xi >a$. In these expressions, $\xi $ has to be substituted from (\ref{coord}). As we see the resulting energy-momentum tensor is non-diagonal. It follows from (\ref{normal}) that the induced metric on the brane is also non-diagonal.

Now we turn to the investigation for the limiting cases of the general formulae for the vacuum energy-momentum tensor. First of all let us consider the near horizon limit, $r\to \alpha $, for a fixed $\theta \neq 0$. In this limit one has $\xi \to 0$ and we can use the results from Ref. \cite{Saha02} for this limit of the Rindler part. As a result we obtain
\begin{eqnarray}
\left\langle T_{0}^{0}\left[ g_{lm},\varphi \right]\right\rangle
^{(b)} &=& -\frac{(D-1)B_0 \sin ^{2(1-D)}(\theta /2)}{4^{D}\pi ^{\frac{D+1}{2}}D \Gamma \left( \frac{D-1}{2}\right) a^{D-1}(\alpha ^2-r^2)\ln ^2(\sqrt{\alpha ^2-r^2}/a)} , \label{ashor1}\\
\left\langle T_{1}^{1}\left[ g_{lm},\varphi \right]\right\rangle
^{(b)} &=&  -\left\langle T_{0}^{0}\left[ g_{lm},\varphi \right]\right\rangle
^{(b)} , \label{ashor2}\\
\left\langle T_{2}^{1}\left[ g_{lm},\varphi \right]\right\rangle
^{(b)} &=& \frac{1}{\alpha } \cot \frac{\theta }{2} \left\langle T_{1}^{1}\left[ g_{lm},\varphi \right]\right\rangle
^{(b)} , \label{ashor3}\\
\left\langle T_{3}^{3}\left[ g_{lm},\varphi \right]\right\rangle
^{(b)} &=& \frac{2\left\langle T_{1}^{1}\left[ g_{lm},\varphi \right]\right\rangle
^{(b)} }{(D-1)\ln (\sqrt{\alpha ^2-r^2}/a)} , \label{ashor4}
\end{eqnarray}
where
\begin{equation}
B_0=\int _{0}^{\infty }dy\, y^{D-2}\frac{\bar K_0(y)}{\bar I_0(y)} .
\label{B0}
\end{equation}
As we see the boundary part is divergent at the dS horizon. Recall that near the horizon the purely dS part behaves as $(\alpha ^2-r^2)^{-(D+1)/2}$ and, therefore, in this limit the total vacuum energy-momentum tensor is dominated by this part.

The boundary induced parts (\ref{tikv1})--(\ref{tikv4}) diverge on the boundary, corresponding to the limit $\xi \to a$. In this limit, by taking into account that $\left\langle T_{1}^{1}\left[ g_{{\mathrm{R}}lm},\varphi
_{{\mathrm{R}}}\right]\right\rangle ^{(b)}\sim (\xi /a-1)\left\langle T_{2}^{2}\left[ g_{{\mathrm{R}}lm},\varphi _{{\mathrm{R}}}\right]\right\rangle ^{(b)}$, for $r\neq \alpha $ we can omit the terms containing $\left\langle T_{1}^{1}\left[ g_{{\mathrm{R}}lm},\varphi _{{\mathrm{R}}}\right]\right\rangle ^{(b)}$ and obtain the following relations between the boundary induced components:
\begin{equation}
\left\langle T_{3}^{3}\right\rangle ^{(b)} \sim -\frac{\left\langle T_{0}^{0}\right\rangle ^{(b)}}{D-1}\sim \frac{\alpha ^2\left\langle T_{1}^{1}\right\rangle ^{(b)}}{a^2 \sin ^2\theta }\sim \frac{r\Omega ^2\left\langle T_{2}^{1}\right\rangle ^{(b)}}{\sin \theta (\cos \theta -r/\alpha )}\sim \frac{\Omega ^2\left\langle T_{2}^{2}\right\rangle ^{(b)}}{(\cos \theta -r/\alpha )^2}, \label{asneara}
\end{equation}
where $\left\langle T_{0}^{0}\right\rangle ^{(b)}\sim (\xi -a)^{1-D}$ and we can substitute in the coefficients of these relations $\cos \theta =(\alpha /r )(1-\sqrt{\alpha ^2-r^2}/a)$. Near the point $(r,\theta )=(\alpha ,0)$, where the boundary touches the horizon, the horizon and boundary divergencies are mixed: in the coefficients of Eqs. (\ref{tikv1})--(\ref{tikv4}) one has $\Omega \to 0$ and from the Rindler parts factors $(\xi -a)^{1-D}$ come.

In the discussion above we have considered the vacuum energy-momentum tensor of the bulk. For a scalar field on manifolds with boundaries in addition to the bulk part the energy-momentum tensor contains a contribution located on the boundary. For arbitrary bulk and boundary geometries the expression of the surface energy-momentum tensor is given in Ref. \cite{Saha03emt}. Special cases of flat, spherical and cylindrical boundaries in the Minkowski background are considered in Refs. \cite{Rome02,Saha01sp,Rome01cyl}. In the case of a conformally coupled scalar field the transformation formula for the surface energy-momentum tensor under the conformal rescaling of the metric is the same as that for the volume part. For our problem in this paper, the surface energy-momentum tensor is obtained from the corresponding Rindler counterpart by a way similar to that described above. The expression for the latter is given in Ref. \cite{Saha03emt}.

\section{Conclusion} \label{secconc}

In the present paper we have investigated the Casimir densities in dS spacetime for a
conformally coupled massless scalar field which satisfies the
Robin boundary condition (\ref{boundcond}) on a hypersurface described by equation (\ref{hypersurf}). The coefficients in the boundary condition are given by relations (\ref{relcoef1}) with constants $A_{{\mathrm{R}}}$ and $B_{{\mathrm{R}}}$ and, in general, depend on the point of the hypersurface. The latter is the conformal image of the flat boundary moving by uniform proper acceleration in the Minkowski spacetime. We have assumed that the field in dS spacetime is in the state conformally related to the Fulling-Rindler vacuum. The energy-momentum tensor in  dS spacetime is generated from the corresponding results in the Rindler spacetime by using the standard formula for the energy-momentum tensors in conformally related problems in combination with the appropriate coordinate transformation. The Rindler energy-momentum tensor is taken from Ref. \cite{Saha02}, where the general case of the curvature coupling parameter is considered. The VEV of the energy-momentum tensor for a brane in dS spacetime consists of two parts given in Eq. (\ref{TikdS1}). The first one corresponds to the purely dS contribution when the boundary is absent. It is determined by formula (\ref{Tik0dSst}), where the second term on the right is due to the trace anomaly and is zero for odd spacetime dimensions. The second part in the vacuum energy-momentum tensor is due to the imposition of boundary conditions on the fluctuating quantum field. The corresponding components are related to the vacuum energy-momentum tensor in the Rindler spacetime by Eqs. (\ref{tikv1})--(\ref{tikv4}) and the Rindler tensor in the region $\xi >a$ is given by formulae (\ref{TikRb})--(\ref{Figz1}). The results for the region $\xi <a$ are obtained from these formulae by replacements $I_{\omega }\to K_{\omega }$, $K_{\omega }\to I_{\omega }$. Unlike to the purely dS part, the boundary induced part of the energy-momentum tensor is non-diagonal and depends on both dS static coordinates $r$ and $\theta $. At the dS horizon both parts in the vacuum energy-momentum tensor diverge, with the leading divergence $(\alpha ^2-r^2)^{-(D+1)/2}$, coming from the purely dS part. Another type of divergence arise on the brane, where the boundary induced part dominates. Near the points where the brane touches the horizon, the divergences are mixed and are stronger. Note that in this paper we have considered vacuum densities which are finite away from the brane and dS horizon. As it has been mentioned in Ref. \cite{Grah02}, the same results will be obtained in the model where instead of externally imposed boundary condition the fluctuating field is coupled to a smooth background potential that implements the boundary condition in a certain limit.

\section*{Acknowledgement }

The work of AAS was supported in part by the Armenian
Ministry of Education and Science (Grant No.~0887).

\end{document}